\documentclass[english,11pt]{article}
\usepackage{geometry}
\usepackage{float}
\usepackage{pgf,tikz}
\usepackage{mathrsfs}
\usetikzlibrary{arrows}
\usepackage{mathrsfs,bm}
\usepackage{amsmath}
\usepackage{amssymb}
\usepackage{fancyhdr}
\usepackage{epsfig}
\usepackage{amsthm}
\usepackage{mathrsfs}
\usepackage{amsfonts}
\usepackage{mathrsfs}
\usepackage{enumerate}
\usepackage{subfigure}
\usepackage{setspace}
\usepackage{showlabels}
\usepackage{mathtools}
\usepackage{tikz}
\usepackage{tikz-3dplot} 
\usepackage{pgfplots}

\usepackage{amsmath,amssymb}
\usepackage{caption}
\usepackage{color}
\usepackage{dcolumn}
\usepackage{bm}
\usepackage{tikz}
\usepackage{tikz-3dplot} 
\usepackage{pgfplots}
\pgfplotsset{compat=1.11}
\usepackage{algorithm}
\usepackage[noend]{algpseudocode}
\usepackage{tikz}
\usetikzlibrary{bayesnet}
\usepackage{amssymb}
\usepackage{caption}
\usepackage{color}
\usepackage{dcolumn}
\usepackage{etoolbox}

\usepackage{amsmath}
\usepackage[noend]{algpseudocode}

\usepackage[utf8]{inputenc}

\usepackage[final]{hyperref} 
\usepackage{setspace}

\usepackage[]{biblatex}
 
\hypersetup{
	colorlinks=true,       
	linkcolor=blue,        
	citecolor=blue,        
	filecolor=magenta,     
	urlcolor=blue         
}
\usepackage{breqn}
\title{Bayesian Model Selection on Random Networks}
\author{Marios Papamichalis \thanks{Purdue University, Postdoctoral Researcher at the Statistics Department, mpapamic@purdue.edu.} }

\begin{document}

\maketitle
\begin{abstract}
A general Bayesian framework for model selection on random network models regarding their features is considered. The goal is to develop
a principle Bayesian model selection approach to compare different fittable,
not necessarily nested, models for inference on those network realisations.
The criterion for random network models regarding the comparison is formulated via Bayes factors and penalizing using the most widely used loss
functions. Parametrizations are different in different spaces. To overcome
this problem we incorporate and encode different aspects of complexities in
terms of observable spaces. Thus, given a range of values for a feature, network realisations are extracted. The proposed principle approach is based
on finding random network models such that a reasonable trade off between
the interested feature and the complexity of the model is preserved, avoiding
over-fitting problems. \\


\end{abstract}

\newpage

\section{Introduction}

In this paper, we propose a general approach for generating and comparing
random network models regarding their features. Given a random network
model, we consider the joint distribution between the parameters and features of the model
we are interested in and generate respective models that follows this distribution.
By formulating the joint distribution we can find a part of distribution which captures a given desired distribution for either the parameters or the features. Different random network
models capture differently parameters or features, sometimes not at all. \\

The researcher might be interested in observing a feature of the random network model (e.g. degree distribution or transitivity) and
there is no obvious model that encodes the features you are interested in making it impossible or difficult to parametrize a model in terms of those features. He can only use models that
understand, interpret, implement and can actually fit, because of computational,
logistic tools, and carry out a secondary analysis if this random network model is
appropriate for describing this feature. For example, SBM does not parametrize
average path length but is pretty flexible model and you have the
means to fit. We want to answer the following questions: Which model, that has
the resources to fit, can capture the feature better? How limited is our description? Maybe the model is not powerful enough to say much things to represent
the feature. We want a criteria that says if some features are well behaving. For
instance, a researcher might be interested in degree distribution and
suppose that he can only fit an Erd{\"o}s-R{\'e}nyi or a SBM model with certain parameters, degree density and number of blocks, respectively. Which
one describes better a certain degree distribution?\\

Our research relates to the literature on intrinsic Bayes factor, model selection on parameters and features on random network models and the literature on
informative prior elicitation. In \cite{Speed} the authors are dealing with the elicitation of informative priors on
graph space that encode parameters and features. In \cite{Yan} the author proposes a Bayesian framework for choosing the number
of blocks as well as comparing it to the more elaborate degree-corrected
block models, ultimately leading to a universal model selection framework
capable of comparing multiple modeling combinations. In \cite{bickel} they consider an approach based on the log likelihood ratio statistic
and analyze its asymptotic properties under model misspecification in order to solve problems concerning estimating the latent node labels and the
model parameters than the issue of choosing the number of blocks. They
show the limiting distribution of the statistic in the case of underfitting is
normal and obtain its convergence rate in the case of overfitting. In \cite{Li}, the authors do not assume only community based penalties but
any specific model for the network. They make a more general structural assumption of a model being approximately low rank, which holds for most popular network models. This is their limitation in contrast with our method. In \cite{Chen} paper, the authors are limited in model selection under block models
and its variants. They focus on a generic idea of network cross-validation.
Cross-validation is a very popular and appealing method in many model selection problems. The adaptation to network data is usually through a node
splitting procedure and has been considered by \cite{Edo}, \cite{Neville}, among others. In \cite{Yan, Schwarz} the author develop a Bayesian model selection criterion for stochastic block models which is inspired by BIC. In \cite{exp} the contribution of this paper is the development of a fully Bayesian
model selection method based on a reversible jump Markov chain Monte
Carlo algorithm which estimates the posterior probability for each competing model. Conceptually, the closest to our research framework is that of \cite{Ivona} where the
authors design and implement MCMC algorithms for computing the maximum likelihood for four popular models: a power-law random graph model, a preferential attachment model, a small-world model, and a uniform random graph model. However there limitation is that their method is confined
by those four models. Moreover, they do not use a loss function as in this
we do in this paper.\\

Here we propose a
methodology for generating and comparing random network models in such a
way that the top model tend to produce posteriors that preserve simplicity and information for a certain feature when compared to posteriors obtained from
models less favored by the comparison. The selection of random network models
is obtained by decision theory approach. We make the case that the selection of
random networks models implied by our approach are reasonably consistent with
selection implied both by the most widely used loss functions functions for estimation and prediction.\\

The approach we propose is creating a joint distribution for a random network
model which encapsulate both the parameters and features of the network. For this reason we use subsampling procedures by producing
network instances for random network models and computing their joint feature
distribution. Next, we are based on Bayesian model selection concepts and tools and we compare
these models using a Bayesian model selection approach. The criterium for random network models regarding the comparison is formulated via Bayes factors of
the most widely used loss functions. The rationale behind the proposed approach
is to find random network models such that preserve the best possible trade of the
information provided for the feature and the complexity of the model from
the original data generating mechanism avoiding the problem of over-fitting.Thus, computational and Bayesian statistics enable us to
generate and compare, in a principled way, random network models.\\

The paper proceeds as follows: In Section 2, we describe settings of the
problem, we formulate them and give notation and definitions of networks and
random networks. Furthermore, we present, briefly, subsampling and Bayes factors tools that are useful in the next sections. Then, in section 3, we focus in our
main purpose of this paper which is how we use and compare random network
models by using decision theory in order to preserve the needed amount of information for a certain given feature or number of features required by Bayes factors.
Conceptually and computationally, our methodology is presented. In section 4,
for many random network models data analysis that gives experimental results involving decision theory is conducted showing the results of our approach. Finally,
in section 5, we present with more details the limitation of the method and future
work involving overlapping research areas.

\section{Preliminaries}

\subsection{Random Networks}

We define a network as a pair $G = (V, E)$, where $V$ denotes the set of nodes, and
$E$ the set of edges $E \in V \times V$ . We denote by $A_G$ the adjacency matrix of $G$.
Let $N$ denote $\mid V \mid$. A network is called simple if at most one edge exists between
each pair of nodes and no self-loops are allowed. A network is called undirected if
the corresponding adjacency matrix is symmetric. A random network (or random
graph model) is a probability model on the space of adjacency matrices. In this
paper we consider random network models in the space of simple undirected
networks, i.e., a distribution on the space of binary symmetric adjacency matrices.
For the sake of simplicity, we are using the same symbol ($G$) for a network and a
random network. We use $G(\omega)$ to denote a realization from the random network
model. The simplest example of random network is Erd{\"o}s-R{\'e}nyi model, where each possible edge in the graph is included with a constant probability $p$. \\

\subsection{Complexity of a Random network}
\begin{figure}
\begin{center}

\tdplotsetmaincoords{70}{110}
\begin{tikzpicture}[scale=4,tdplot_main_coords]
\draw[thick,->] (0,0,0) -- (1,0,0) node[anchor=north east]{$x$};
\draw[thick,->] (0,0,0) -- (0,1,0) node[anchor=north west]{$y$};
\draw[thick,->] (0,0,0) -- (0,0,1) node[anchor=south]{$z$};
\tdplotsetcoord{P}{.8}{50}{70}

\draw[color=red] (0,0,0) -- (0.78,0.78,0.78) node[anchor=west]{$(a_1,a_2,a_3)$};

\draw[dashed, color=red] -- (Pxy);
\draw[dashed, color=red](P) -- (Pxy);
\tdplotsetthetaplanecoords{70}
\draw[tdplot_rotated_coords,color=blue,thick,->] (0,0,0)
-- (.2,0,0) node[anchor=east]{$x^{'}$};
\draw[tdplot_rotated_coords,color=blue,thick,->] (0,0,0)
-- (0,.2,0) node[anchor=north]{$y^{'}$};
\draw[tdplot_rotated_coords,color=blue,thick,->] (0,0,0)
-- (0,0,.2) node[anchor=west]{$z^{'}$};
\end{tikzpicture}
\end{center}
\caption{Penalizing in the space of parameters is complex as we can see. $a_1$ denotes the community structure. $a_2$ denotes the degree distribution and $a_3$ the network subcounts}
\end{figure}
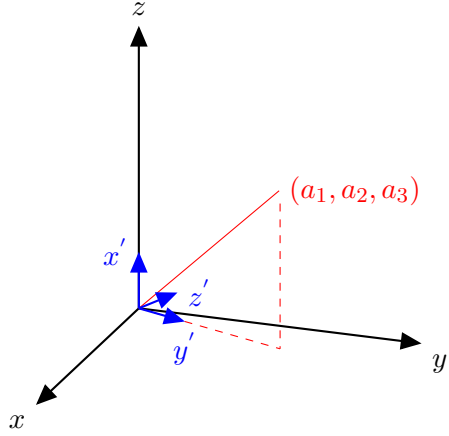

A mathematical framework has been proposed for analyzing Random Network
Models and for characterizing their complexity. Such framework allows the study
of several network properties or features (link density, clustering coefficient, degree distribution, connectivity), and their relationship with the random network
model complexity. For doing so, different entropy measures have been evaluated
and their relationship has been assessed. The sample degree distribution entropy
has shown to be correlated with the random network model entropy, providing
a practical measurable indicator of complexity in real networks. Generally, link
density, clustering coefficient, degree distributions and connectivity are computed
and comparatively analyzed in order to illustrate their relationship with random
network model complexity (Figure 1). The fundamental simulations have been performed
by imposing varying values of the link density and triangle density parameters in
ERGMs.

\subsection{Bayes Factor}

One of the central quantities in Bayesian learning is the evidence (or marginal
likelihood) , the probability of the data given the model $P(D \mid M_i)$ computed as
the integral over the parameters w of the likelihood times the prior. The evidence
is related to the probability of the model, $P(M_i \mid D)$ through Bayes rule:
\begin{equation}
P(D \mid M_i) = \int P(D \mid w, M_i)P(w \mid M_i)dw, P(M_i \mid D) = P(D \mid M_i)P(M_i)P(D)
\end{equation}
where it is not uncommon that the prior on models $P(M_i)$ is flat, such that $P(M_i \mid D)$ is proportional to the evidence. It is typically impossible to compute analytically. However, the model evidence is crucial for Bayesian model selection since
it allows us to make statements about posterior model probabilities. The evidence
discourages overcomplex models, and can be used to select the most probable
model. It is also possible to understand how the evidence discourages overcomplex models and therefore embodies Occam's Razor by using the following interpretation. The evidence is the probability that if you randomly selected parameter
values from your model class, you would generate data set $D$. Models that are too
simple will be very unlikely to generate that particular data set, whereas models
that are too complex can generate many possible data sets, so again, they are unlikely to generate that particular data set at random. To the point, on computation, we will evaluate the different likelihood integrals in the Bayesian setting with a large-scale Monte Carlo procedure in almost any case of practical interest.

\subsection{Intrinsic Expected Losses}

As we mentioned  in the introduction, many
methods for model selection between random networks have been developed. All
of them are penalizing based on one important attribute of the network which
tries to measure the complexity of the network, as it is described in the previous
subsection. Moreover, all of them select models from the same family of models
e.g. are able to select which SBM fits better the data without overfitting them.
Non of the mentioned methods are able to perform a principled universal model
selection between two different random network models (e.g. compare a certain
Barabasi-Albert model with a certain SBM). \\

To address the natural question of which model is best for a particular data
set, we propose a model selection criterion for graph models and present a more flexible principle model selection criterion inspired by Bayesian Decision Theory
penalizing with Expected Losses of one or more specific features. The Bayes decision is simply the hypothesis with the larger posterior probability. The posterior
expected loss of two models $M_1$ and $M_2$ are $K_1 \times P(M_1 \mid D)$ and $K_2 \times P(M_2 \mid D)$, respectively. The Bayes decision is again treat corresponding to the smallest posterior expected loss. In this Bayes test, the null hypothesis is rejected when:
\begin{equation}
\frac{K_1}{K_2}>\frac{P(M_1 \mid D)}{P(M_2 \mid D)}
\end{equation}
In statistics and decision theory a loss function or cost function is a function
that maps an event or values of one or more variables onto a real number intuitively
representing some ”cost” associated with the event. Simply, we formalize good
and bad results with a loss function. The loss function determines the penalty for
deciding how well a model is behaving in terms of the feature which is going to
penalize it. Some examples involve: i) 0-1 loss function where $L(\hat{y}, y) = I(\hat{y} \neq y)$ and $I$ is the indicator
function. ii) Quadratic Loss function where for a scalar parameter $\theta$, a decision function
whose output $\hat{\theta}$ is an estimate of $\theta$, and a quadratic loss function $L(\theta, \hat{\theta}) = (\theta - \hat{\theta})$ and iii) absolute Loss function where for a scalar parameter $\theta$, a decision function whose
output $\hat{\theta}$ is an estimate of $\theta$, and a quadratic loss function $L(\theta, \hat{\theta}) =\mid \theta -\hat{\theta} \mid$.\\

Since the choice of a particular loss function strongly influences the resulting
inference, it seems necessary to rely on intrinsic losses when no information is
available about the utility function of the decision-maker, rather than to call for
classical losses like the squared error loss. Since this setting is quite similar to
the derivation of non-informative priors in Bayesian analysis, we first recall the
conditions of this derivation and deduce from these conditions some requirements
on the intrinsic losses. For that reason intrinsic loss functions could be used.

\subsection{Extracting Random graph information}

We would like to define a prior such that, when sampling from the prior, we generate reasonable random networks. Here, reasonable is taken to mean that the networks should respect graph-theoretical properties that have been inferred by
network properties. A general form of informative log linear distribution over
graph is proposed by \cite{Speed}. The random graph prior is defined as a distribution:
\begin{equation}
P(G) \propto
 exp(\lambda \sum_i w_i f_i(G))
\end{equation}
,in which $\lambda$ is a strength parameter and the parameters $w_i$ tune the relative
strengths of the individual concordance functions that capture several graph-theoretical
properties of random network realization properties. Some examples of concordance functions are: individual edges, controlling the in-degree of graphs, higher-level Network features, degree distributions, priors on individual edges and priors on degrees counts. \\

\subsection{Exchangeability and Concensus Monte Carlo}

For estimating features from exchangeable models which are not straightforward to relate to the parametrization at hand, we extract the information about the feature included in the empirical graphon through the corresponding SBM of the model (each node/edge/measure-exchangeable model (\cite{Lovasz}, \cite{Broderick},
\cite{Caron},
\cite{Crane}) can be divided into parts in line with empirical graphon and described by it). This task is high dimensional (NP-complete) and in order to make it scalable we divide the model into $n \times n$ cells of 10 nodes each. When exchangeable or more generally model exchangeable parametric objects, through empirical graphon, are available then for estimating features from exchangeable models which are not analytically parametrized e.g. random networks producing diameters, we extract the information through \cite{Speed}. Let $f(G)$ be a real-valued function on graphs that
is increasing in the degree to which graph $G$ agrees with prior
beliefs (a ‘‘concordance function’’). For potentially multiple concordance functions $f_i(G)$, we suggest a log-linear
network prior of the form:

\begin{equation}
P(G) \propto \exp\{\lambda \sum_{i=1}^{n^2} w_i f_i(G)\}=\prod_{i=1}^{n^2} \exp \{ \lambda w_i f_i(G) \}
\end{equation}

, where the $\exp\{\lambda w_i f_i(G)\}$ is one cell of the model which describes a submodel (e.g one  can be scale free distribution and the other follow a constant distribution).\\ 

Then, we merge all the networks into one each time, extracting the feature (parameter) (\cite{Ying}). For this, we use concensus monte carlo (\cite{Scott}), in order to merge the observables for each feature. Finally, we use all those merged observables using intrinsic loss functions for each feature. Obviously, this approach is computationally expensive, though it is scalable, due to model division in $n \times n$ squared cells (number of cores needed in Map-Reduce framework). Depending on the number of cores used we can reach complex networks up to 10.000 nodes.

\section{Methodology}

\begin{center}
\begin{figure}
\begin{center}
    \begin{tabular}{cc}
%
%
%


\begin{tikzpicture}


\node[latent]            (t) {$\theta$};
\node [latent, above=1.3cm of t] (d){$f(\theta$)};
\node[latent,right=0.5cm of t]   (q){$p(\mathcal{G}_\theta \mid \theta)$};
\node[latent,right=0.5cm of q]   (p){$p(a_{\mathcal{G}_\theta} \mid \theta)$};

\edge{d}{t};
\edge{t}{q};
\edge{q}{p};

  \plate {} {(t)} {$$} ;

\end{tikzpicture}

    \end{tabular}
  \end{center}
\caption{Procedure for producing observables.}
\end{figure}
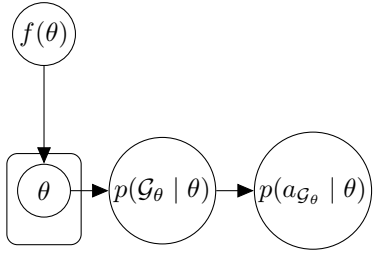
\end{center}

\subsection{General Concepts}
Suppose we have an observed feature (e.g. centrality) and we are given two different network models (e.g. SBM and Erd{\"o}s-R{\'e}nyi). What is the probability of observing the
observed feature  given the first model (e.g. SBM) compared with the probability of observing it given the second model (e.g. Erd{\"o}s-R{\'e}nyi model)? More specifically:
\begin{equation}
BF =\frac{Pr(D_a \mid M_1)}{Pr(D_a \mid M_2)}
\end{equation}
Here $D_a$ are networks produced from the concordance function of centrality
for a specific regimes of centralities, $M_1$= Centralities produced by a SBM with
parameters $K$, $z$ and $M_2$= Centralities produced by an Erd{\"o}s-R{\'e}nyi with parameters $\theta$. $M_1$
and $M_2$ are produced in the space of observables. We sample SBM and Erd{\"o}s-R{\'e}nyi models and for each realization we produce networks. For those networks we calculate
their centralities. Given those networks we produce the distribution of centralities
for this model (SBM or Erd{\"o}s-R{\'e}nyi respectively). Model comes from prior predictive
(Figure 2) in random variable of centrality $a$: $Pr(a \mid K)$ for SBM and $Pr(a \mid \theta)$
for Erd{\"o}s-R{\'e}nyi. If the observed feature is discrete (like diameter which is the maximum distance between nodes) then we are fine. If the observed feature is continuous we can
compute posterior probability and use a smoothing method like a kernel density
function to get the density.
\begin{center}
\begin{figure}
\begin{center}
    \begin{tabular}{cc}
%
%
%


\begin{tikzpicture}


\node[latent]            (t) {$\hat{\alpha}$};
\node [latent, below=0.5cm of t] (d){$p(\hat{\alpha})$)};
\node[latent,below=0.5cm of d]   (w){$G$ with $\hat{\alpha}$};

\edge{t}{d};
\edge{d}{w};

\end{tikzpicture}

    \end{tabular}
  \end{center}
\caption{Practitioner's tool for tweaking priors.}
\end{figure}
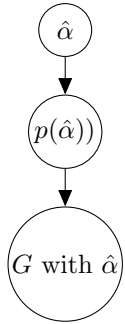
\end{center}

Another thing we want to achieve is to provide a tool (figure 3) to formulate a question
in terms of ranges, which can be useful for prior elicitation. In networks is very
hard for practitioner to tell you information in terms of parameters (uncertainty
in terms of parameters). Practitioners that are interested in specific ranges of features and once they have those ranges they can set candidates for the priors and see
which one is more reasonable-better for those ranges. This tool is powerful in
terms of flexibility way for looking of ranges of features. So we are interested in
answering how the prior of the network should be? That is very hard. Networks
are very complicated objects and hard to be interpreted. Practitioners can look at
the distribution of low dimensional features, get intervals that can be used in ways
to calibrate your prior or to compare different models which could serve as a prior.
We can achieve this by calculating the Bayes Factor:
\begin{equation}
BF =\frac{Pr(D \mid M_1)}{Pr(D \mid M_2)}
\end{equation}
Here $D$ are general network data, $M_1$= SBM with parameters $K$, $z$ given $\hat{\alpha}$ and
$M_2$=Erd{\"o}s-R{\'e}nyi with $\theta$ given $\hat{\alpha}
$. $M_1$ and $M_2$ are produced in the space of observables. We sample SBM and Erd{\"o}s-R{\'e}nyi models and for each realization we produce networks.\\

For those networks we calculate their centralities. Given those networks we produce the distribution of centralities for this model (SBM or Erd{\"o}s-R{\'e}nyi respectively).
Model comes from: $Pr(K, \hat{a})$ in SBM and $Pr(\theta, \hat{a})$ in Erd{\"o}s-R{\'e}nyi.

\subsection{Model Selection of networks with respect to more than
one feature.}

Which model is better to fit to network data $D$? Model $M_1$ or model $M_2$? Until
now we use the Bayesian factor test for constructing a principle Bayesian model
selection test. How do we combine three universal features such as community
structure, degree distribution and network subcounts in a model selection test in
a plausible principle way that makes sense? The penalty of the one might dominate the other and we might encounter problems where e.g. community structure is much more dominant than degree distribution or the opposite because of their
values (large, small).\\

The Left Hand Side in (8), (9) and (10) is a number which absorbs
(cancel out) any scaling problem from (11). You do not need to scale the axis every time to find the right portion of inclusion of each parameter’s penalty. Q.E.L.
stands for the quadratic expected loss function in observable space. $D_{M_{1,2}}$
are
the networks which are produced by the models $M_{1,2}$, respectively, and which
we want to compare. We extract their features and compare them (Number of
blocks, entropy of Degree distribution and number of Subgraph counts) which the
corresponding features of the network data $D$. Which model could be fitted best
regarding those three universal features? With this approach we do not encounter
any problem regarding a dominance feature (scaling penalties) as in (5) which is
derived if we directly use all the penalties from \cite{Berger1}
\begin{equation}
\frac{Q.E.L(D_{M_1},D)_{Blocks}}{Q.E.L(D_{M_2},D)_{Blocks}}><\frac{P(M_1 \mid D)}{P(M_2 \mid D}
\end{equation}

\begin{equation}
\frac{Q.E.L(D_{M_1},D)_{D.D.}}{Q.E.L(D_{M_2},D)_{D.D.}}><\frac{P(M_1 \mid D)}{P(M_2 \mid D}
\end{equation}

\begin{equation}
\frac{Q.E.L(D_{M_1},D)_{Motifs}}{Q.E.L(D_{M_2},D)_{Motifs}}><\frac{P(M_1 \mid D)}{P(M_2 \mid D}
\end{equation}

Summing the above 3 equations and divide by 3 we have:

\begin{equation}
\frac{\frac{Q.E.L(D_{M_1},D)_{Blocks}}{Q.E.L(D_{M_2},D)_{Blocks}}+\frac{Q.E.L(D_{M_1},D)_{D.D.}}{Q.E.L(D_{M_2},D)_{D.D.}}+\frac{Q.E.L(D_{M_1},D)_{Motifs}}{Q.E.L(D_{M_2},D)_{Motifs}}}{3}><\frac{P(M_1 \mid D)}{P(M_2 \mid D}
\end{equation}

\section{Simulation Studies}
We will perform three types of simulations. In all of them we will use Barabasi-Albert model and Stochastic Block Model. For the Barabasi-Albert model we
will produce networks given the power law degree distribution sampling. For the
Stochastic Block Model we will use \cite{latouche} method. The three simulation setups are
described below. The number of samples is N=100.
\begin{table}
\centering
\fbox{%
\begin{tabular}{| l   l  l   |}
\hline
Random Network Produced & Read Data Parameters & Tested Parameters \\
\hline            
Barabasi Albert & $\alpha=3.2$ & $P(\alpha=[2.9, 3.1] \mid D)$\\
\hline
Stochastic Block Model & $K$=10 blocks & $P(K=9\mid D)$\\
\hline
\end{tabular}}
\caption{Real networks are constructed through Barabasi-Albert and Stochastic Block Model}
\end{table}
\begin{table}
\centering
\fbox{%
\begin{tabular}{| l     |}
\hline
Local Feature  \\
\hline            
Power Law\\
\hline
SBM\\
\hline
Degree distribution through entropy\\
\hline     
\end{tabular}}
\caption{Feature according to which we select a model}
\end{table}
\begin{table}
\centering
\fbox{%
\begin{tabular}{| l   l   l  |}
\hline
Loss Function& Expression & Type of Inference \\
\hline            
Quadratic Loss & $L(\theta,\hat{\theta})=(\theta-\hat{\theta})^2$ &  point estimation\\
\hline
Absolute Loss & $L(\theta,\hat{\theta})=\mid\theta-\hat{\theta}\mid$ & point estimation\\
\hline
\end{tabular}}
\caption{Loss functions used for penalties}
\end{table}
\begin{center}
\begin{table}

\fbox{%
\begin{tabular}{| l   l   l   l   |}
\hline
Real Param. & L.F. Ratio & Tested Parameters Probabilities  & Local Feature\\
\hline            
$\alpha$ &  Q.L.=1493 & $P(\alpha=[2.9,3.1 \mid D)=0.67, P(K=9)=0.010$ &Power Law\\
\hline
$\alpha$ & A.L=0.0328 & $P(\alpha=[2.9,3.1 \mid D)=0.67, P(K=9)=0.010$ &Power Law\\
\hline
$K$ & Q.L.=31.082 & $P(\alpha=[2.9,3.1 \mid D)=0.03, P(K=9)=0.37$  &SBM\\
\hline
$K$ & A.L.=8.0287 & $P(\alpha=[2.9,3.1 \mid D)=0.03, P(K=9)=0.37$ &SBM\\
\hline
$\alpha=3.2$ & Q.L.=0.8685 &$P(\alpha=[2.9,3.1 \mid D)=0.02, P(K=9)=0.03$  &Power law and D.D.\\
\hline
$\alpha=3.2$ & A.L.=0.241&$P(\alpha=[2.9,3.1 \mid D)=0.02, P(K=9)=0.03$ &Power law and D.D.\\
\hline
\end{tabular}}
\caption{Results of the loss functions}
\end{table}
\end{center}
\subsection{Simulation Set Up}
As we can see from Table 5 when the data resemble more to a power law distribution it much more possible to select models that follow the power law distribution and have similar behaviour (their exponents are very close). On the other hand, when the network data are extracted from a SBM then it is much more likely
to select models with a SBM with number of blocks close to one that the read network date were constructed. The same happened with degree equals to 0.4 but
here the results are more ambiguous that in the other two cases.
As we can observe from the last two rows from in the case of the combination of two loss 
functions, one for power law and one for the degree obviously the power law fraction
 dominates the degree distribution fraction. Still in that case power law distribution is
much more possible to select models that follow the power law distribution and have 
similar behavior. 

\section{Discussion}

Our methodology constitutes an attempt for comparing random network models
regarding local features by using intrinsic Bayesian factor. We investigated how
the following three interact: i)feature that you have phrase your scientific questions ii) features that are formulated by the model you are fitting and iii) features
that can be retrieved more efficiently having sample size you have. For example:
i) scientific question in terms of transitivity ii) you do not have a model that can be
generated in terms of transitivity and iii) the compromise in terms of the feature
what you want to capture and the model you can fit.\\

The need to compare random networks is fundamental to many fields, from the
physical and life sciences to the social, behavioral, and economic sciences. This
need is currently particularly significant in understanding information and incorporating complexity of random networks in network model selection. Although all
model selection methods address the inevitable trade-off between goodness-of-fit
and complexity, the manner in which they measure and penalize model complexity can differ substantially. The main advantage of our method is that it enables the
statistician, for the first time, to compare random network models, both nested and
non-nested and overcome the complexity problem. To this end, we have shown,
through a comparative analysis of several rich and varied examples that the results
obtained suggest that our method is both reasonable and universal.\\

The limitations of the approach in its current form include: i) Prior information of the data have to be incorporated to the complexity of the model in order to
penalize the model, so it is reasonable the model to be related with the data ii) In
multidimensional random network models computational costs might be an issue.\\

Future work includes: i) how can we reduce the cost of our algorithms in terms
of complexity in high dimensional random network model features in the model
selection ii) is it possible to create concordance function for every local feature
regimes? ii) is there another way to penalize models due to features like e.g. modularity, motifs work or use two dimensional entropy of a graphon, as a penalty? iv) is it possible to connect asymptotically (e.g. through von Mises theorem) this approach for node-exchangeable models with the work in \cite{bickel}?

\section*{Acknowledgement}

The author thanks Sim\'on Lunag\'omez from  Department of Mathematics and Statistics at Lancaster University for his valuable comments.

\printbibliography

\end{document}